\documentclass[aps,prl,floatfix,epsfig,twocolumn,showpacs,preprintnumbers]{revtex4}
\usepackage{amssymb}
\usepackage{graphicx}
\usepackage{amsmath}

\begin{document}

\title{Dirac and Weyl Semimetal in $XY$Bi ($X$=Ba, Eu; $Y$=Cu, Ag and Au)}
\author{Yongping Du$^{1}$, Bo Wan$^{1}$, Di Wang$^{1}$, Li Sheng$^{1,2}$,
Chun-Gang Duan$^{3}$ and Xiangang Wan$^{1,2}$}
\thanks{Corresponding author: xgwan@nju.edu.cn}
\affiliation{$^{1}$National Laboratory of Solid State Microstructures and Department of
Physics, Nanjing University, Nanjing 210093, China\\
$^{2}$Collaborative Innovation Center of Advanced Microstructures, Nanjing
University, Nanjing 210093, China\\
$^{3}$Key Laboratory of Polar Materials and Devices, Ministry of Education,
East China Normal University, Shanghai 200062, China}

\begin{abstract}
Weyl and Dirac semimetals recently stimulate intense research activities due
to their novel properties. Combining first-principles calculations and
effective model analysis, we predict that nonmagnetic compounds Ba$Y$Bi ($Y$%
=Au, Ag and Cu) are Dirac semimetals. As for the magnetic compound Eu$Y$Bi,
although the time reversal symmetry is broken, their long-range magnetic
ordering cannot split the Dirac point into pairs of Weyl points. However, we
propose that partially substitute Eu ions by Ba ions will realize the Weyl
semimetal.
\end{abstract}

\date{\today }
\pacs{71.20.-b, 73.20.-r, 71.20.Lp}
\maketitle

Following the discovery of topological insulator (TI) \cite{RE-1,RE-3},
there has been considerable research interest in studying the Weyl semimetal
(WSM), the first metallic topologically nontrivial matter \cite%
{Weyl,Weyl-2,RE-4,Weyl-response}. In WSM, non-degenerate valence and
conduction bands touch at an accidental degeneracy point in a
three-dimensional (3D) Brillouin zone, and its low energy physics is
approximated by the Weyl equation \cite{Weyl,Weyl-2}. Weyl points, the
nondegenerate linear touchings of the bulk bands, always come in pair, and
they are robust due to the protection by the topology of the band structure.
The most remarkable feature of WSM is the Fermi arc surface states \cite%
{Weyl}. Several compounds, including pyrochlore iridates \cite{Weyl}, TI
based heterostructures \cite{Balents-Weyl}, HgCr$_{2}$Se$_{4}$\cite{HgCr2Se4}
and many other systems \cite{Mn-HgTe,Li-Lun-Weyl-2,Weyl-4,Li-Weyl-3} had
been theoretically\ predicted as promising WSMs. While the indication about
realization of WSM have been reported \cite%
{Exper-Weyl-1,Exp-Weyl-2,Exp-Weyl-3}, the presence of the Fermi arc, as the
smoking-gun feature, unfortunately still has not been confirmed.

Same as the WSM, the Dirac semimetal (DSM) is also a 3D analog of graphene 
\cite{DSE-1,DSE-2,DSE-3,DSM-33}. But in contrast with Weyl point, the Dirac
point has four-fold degeneracy, and does not possess the a topological
number, consequently the Dirac point is not robust to against the external
perturbations and usually hard to be realized. Thus the 3D DSM receive much
less attention until the discovery of Na$_{3}$Bi\cite{Na3Bi} and Cd$_{3}$As$%
_{2}$\cite{Cd3As2}. Wang \textit{et al.} find that there is a paired 3D bulk
Dirac points exist on the $k_{z}$ axis of Na$_{3}$Bi \cite{Na3Bi} and Cd$%
_{3} $As$_{2}$\cite{Cd3As2}, and these Dirac points are protected by the
crystal symmetry thus are stable \cite{Na3Bi,Cd3As2}. The theoretical
propose about Na$_{3}$Bi and Cd$_{3}$As$_{2}$ \cite{Na3Bi,Cd3As2} had been
quickly confirmed by the subsequent photoemission measurement \cite{Y.L.
Chen, Cd3As2-2,Hansan,Cava}. This immmediately triggers a new wave of
research to explore the unique properties associated with the 3D Dirac
points in the DSM \cite{Y.L. Chen, Cd3As2-2,Hansan,Cava,Li-SY,DSE-4,DSE-5}.
Unfortunately, Na$_{3}$Bi is not stable in air while arsenic limits the
application of Cd$_{3}$As$_{2}$. Therefore searching new 3D DSM with less
toxic and stability in nature is of both fundamental and technological
importance.

\begin{figure}[tbh]
\center
\includegraphics[scale=0.2]{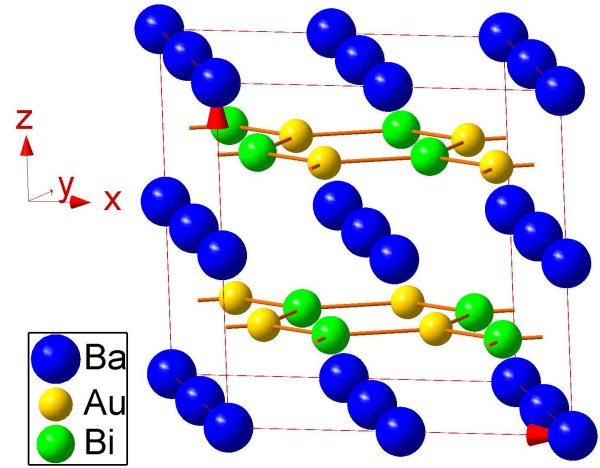}
\caption{Crystal structure of BaAuBi. BaAgBi and BaCuBi have similar
structure.}
\end{figure}

In this paper, based on the density functional theory (DFT) calculations and
effective low energy models, we predict that Ba$Y$Bi ($Y$=Au, Ag and Cu) are
promising 3D Dirac materials. For BaAuBi, the nontrivial topology is due to
the band inversion of the Bi-\textit{p} bonding and antibonding states,
while for the BaAgBi and BaCuBi, the band inversion happens between the
Ag/Cu \textit{s} and Bi \textit{p} orbital. Protected by the $C_{3}$
rotation symmetry, the Dirac points locate along the $\Gamma -A$\ line. The
magnetic configuration in Eu$Y$Bi indeed break the time reversal symmetry,
however cannot split the Dirac point into two Weyl points. We propose that
partially substituting Eu by Ba, i.e. alloy compound Ba$_{x}$Eu$_{(1-x)}$%
Ag(Au)Bi, which could be grown using molecular beam epitaxy (MBE) technique,
is a promising way to realize the WSM.

The electronic band structure calculations have been carried out using the
full potential linearized augmented plane wave method as implemented in
WIEN2K package \cite{WIEN2K}. The modified Becke-Johnson exchange potential
together with local-density approximation for the correlation potential
(MBJLDA) has been used here to obtain accurate band inversion strength and
band order \cite{mBJ}. A $16\times 16\times 7$ mesh is used for the
Brillouin zone integral. Using the second-order variational procedure, we
include the spin-orbital coupling (SOC) interaction.

Ba$Y$Bi ($Y$=Au,Ag,Cu) crystallize in the same hexagonal ZrBeSi type
structure with space group $P6_{3}/mmc\ (D_{6h}^{4})$ \cite{BaXBi structure}%
. The crystal structure of BaAuBi\ is shown as an example in Fig.1, in which
Au and Bi ions form honeycomb lattice layers stacking along \textit{c} axis
and sandwiched by trigonal layers formed by Ba atoms. There are two formula
units in the primitive unit cell, and the six atoms in the unit cell can be
classified as three nonequivalent crystallographic sites: Ba, Au and Bi
according to the symmetry. Ba locate at the 2\textit{a} $(0,0,\frac{1}{2})$,
while Ag and Bi occupy the 2\textit{c} $(\frac{1}{3},\frac{2}{3},\frac{1}{4}%
) $ and 2\textit{d} $(\frac{2}{3},\frac{1}{3},\frac{1}{4})$ sites, there is
no free internal coordinates, and the lattice constants are the only
structural parameter for Ba$Y$Bi lattice. We optimize the lattice parameter
and for all of the three compounds, our numerical lattice constants are in
good agreements with experiments, and the small discrepancy between the
numerical and experimental structure has negligible effect on the electronic
structure. Hence, the following results are obtained based on the
experimental structure, unless stated specifically.

We first calculate the electronic structure of BaAuBi, and show the results
in Fig.2(a). The Ba in BaAuBi is highly ionic, has negligible contribution
around Fermi level. Au-6\textit{s} and 5\textit{d}\ bands mainly located at $%
-$4 to $-$1 eV, and $-$6 to $-$4 eV, respectively. The Bi-6$s$ is basically
located about -11 eV below the Fermi level. The valence and conduction bands
are dominated by the Bi-6\textit{p }bonding and antibonding states. Checking
the wave function, we find that at the $\Gamma $ point the Bi-6\textit{p }%
antibonding state is higher than the Bi-6\textit{p }bonding state, however
at the $A$\ point, the odd-parity state is about 0.545 eV lower in energy
than even-parity state.

\begin{figure}[tbh]
\center\includegraphics[scale=0.5]{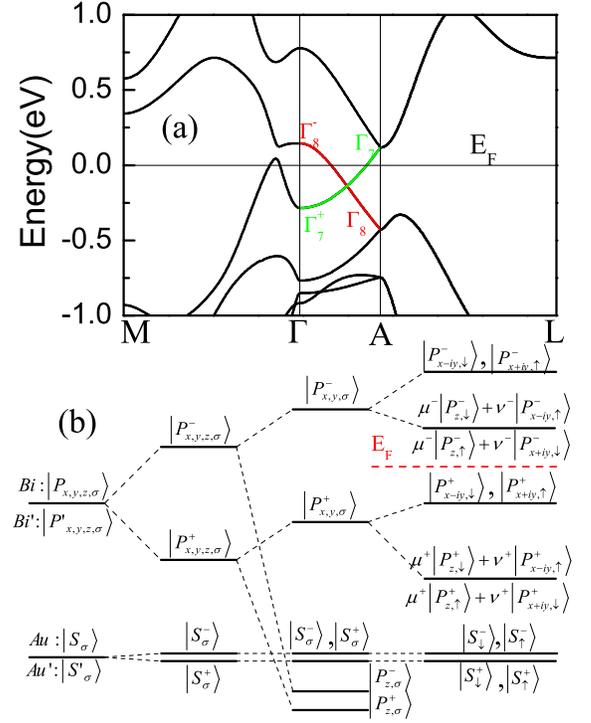}
\caption{(a) Electronic structure of BaAuBi. Green and red line highlights
the different irreducible representation along $\Gamma -A$. (b) Band
evolution near Fermi energy of BaAuBi at $\Gamma $ point, red dashed line
stands for the Fermi energy (see main text for detailed description).}
\end{figure}

In order to understand the mechanism of the band inversion, we illustrate
the band evolution at the $\Gamma $\ point of BaAuBi at Fig.2(b). As discuss
above, the states near Fermi level are primarily contributed by the Bi-6%
\textit{p} orbital, with also the Au-6\textit{s} state. Since the two Bi
atoms (Bi and Bi') in the unit cell are related to each other by the
inversion symmetry, similar with Ref.\cite{HJ Zhang,C-X Liu}, we combine the
Bi-6\textit{p} orbitals to form the hybridized states and label the bonding
and antibonding states as $|P_{x,y,z}^{+}\rangle $\ ($|P_{x,y,z}^{+}\rangle =%
\frac{1}{\sqrt{2}}(|$Bi$,p_{x,y,z}\rangle -|$Bi$^{\prime },p_{x,y,z}\rangle )
$) and $|P_{x,y,z}^{-}\rangle $\ ($|P_{x,y,z}^{-}\rangle =\frac{1}{\sqrt{2}}%
(|$Bi$,p_{x,y,z}\rangle +|$Bi$^{\prime },p_{x,y,z}\rangle )$)\ respectively,
where the superscripts $+/-$ denote the parity of the corresponding states.
According to the point group symmetry, the $p_{z}$ orbital split from the $%
p_{x}$ and $p_{y}$ orbitals while the latter two still degenerate as shown
in the Fig.2(b). Finally, we consider the effect of SOC. The $%
|P_{x+iy,\uparrow }^{+}\rangle $ and $|P_{x-iy,\downarrow }^{+}\rangle $\
states are pushed up by the SOC, while the $|P_{z,\uparrow }^{-}\rangle $ ($%
|P_{z,\downarrow }^{-}\rangle $) will mix with $|P_{x+iy,\downarrow
}^{-}\rangle $ ($|P_{x-iy,\uparrow }^{-}\rangle $), consequently the bonding
($|P_{x+iy,\uparrow }^{+}\rangle $ and $|P_{x-iy,\downarrow }^{+}\rangle $)
and antibonding states ($\mu ^{-}|P_{z,\uparrow }^{-}\rangle +\upsilon
^{-}|P_{x+iy,\downarrow }^{-}\rangle $ and $\mu ^{-}|P_{z,\downarrow
}^{-}\rangle +\upsilon ^{-}|P_{x-iy,\uparrow }^{-}\rangle $) are close to
each other at the $\Gamma $\ point, and the band inversion occurs at the $A$
point as shown in Fig.2.

Along $\Gamma -A$ line the C$_{3}$ symmetry is reserved, by the symmetry
analysis the two relevant bands along this line belong to different
representations ($\Gamma _{7}$ and $\Gamma _{8}$\ as shown in Fig.2(a). See
the appendix for the detail). Thus the hybridization between these bands is
strictly forbidden, which results in the protected band crossing as shown in
Fig.2(a). The Dirac point is located slightly below the Fermi level as shown
in Fig.2(a), and the novel properties associated with DSM\ can be observed
in BaAuBi.

Since the topological nature is determined by the $\Gamma _{7}$ and $\Gamma
_{8}$ bands, based on the projection-operator method (see the appendix), we
build the effective Hamiltonian by using the four relevant states as bases
(in the order of $|P_{x+iy,\uparrow }^{+}\rangle ,\ \mu ^{-}|P_{z,\downarrow
}^{-}\rangle +\upsilon ^{-}|P_{x-iy,\uparrow }^{-}\rangle ,\
|P_{x-iy,\downarrow }^{+}\rangle ,\ \mu ^{-}|P_{z,\uparrow }^{-}\rangle
+\upsilon ^{-}|P_{x+iy,\downarrow }^{-}\rangle $) at $\Gamma $ point. We
neglect all of other states, since they are far from the Fermi level and not
involve into the band inversion, and the Hamiltonian can be written as:

\begin{equation*}
H_{eff}=\epsilon _{0}(\mathbf{k})+\left( 
\begin{matrix}
M(\mathbf{k}) & A(\mathbf{k})k_{+} & 0 & Bk_{z}k_{+}^{2} \\ 
A(\mathbf{k})k_{-} & -M(\mathbf{k}) & -Bk_{z}k_{+}^{2} & 0 \\ 
0 & -Bk_{z}k_{-}^{2} & M(\mathbf{k}) & A(\mathbf{k})k_{-} \\ 
Bk_{z}k_{-}^{2} & 0 & A(\mathbf{k})k_{+} & -M(\mathbf{k})%
\end{matrix}%
\right)
\end{equation*}

where $\epsilon _{0}(\mathbf{k}%
)=C_{0}+C_{1}k_{z}^{2}+C_{2}(k_{x}^{2}+k_{y}^{2})$, $M(\mathbf{k}%
)=M_{0}-M_{1}k_{z}^{2}-M_{2}(k_{x}^{2}+k_{y}^{2})$ ,$A(\mathbf{k}%
)=A_{0}+A_{1}k_{z}^{2}+A_{2}(k_{x}^{2}+k_{y}^{2})$ and $k_{\pm }=k_{x}\pm
ik_{y}$. The parameters in the above formula are material dependent, and by
fitting the DFT calculated band dispersion, we obtain $C_{0}=-0.06978eV$, $%
C_{1}=-0.34038eV\cdot \mathring{A}^{2}$, $C_{2}=2.25eV\cdot \mathring{A}^{2}$%
, $M_{0}=-0.21537eV$, $M_{1}=-1.9523eV\cdot \mathring{A}^{2}$, $%
M_{2}=-7.9507eV\cdot \mathring{A}^{2}$, and $A_{0}=1.3668eV\cdot \mathring{A}
$. Solving the above eigenvalue problem, we obtain\textbf{\ }$E(\mathbf{k}%
)=\epsilon _{0}(\mathbf{k})\pm \sqrt{M(\mathbf{k})^{2}+A(\mathbf{k}%
)^{2}k_{+}k_{-}+|B|^{2}k_{z}^{2}k_{+}^{2}k_{-}^{2}}$\textbf{, }and at $%
\mathbf{k}_{c}=(0,0,\pm \sqrt{\frac{M_{0}}{M_{1}}}),$\textbf{\ }we get the
gapless solutions. In the vicinity of $\mathbf{k}_{c}$\ and neglect the
high-order terms, $E(\mathbf{k}^{^{\prime }})$ would be equal to $\epsilon
_{0}(\mathbf{k}^{^{\prime }})\pm \sqrt{4M_{1}^{2}\mathbf{k}_{c}^{2}\delta
k_{z}^{2}+A^{2}(\mathbf{k}_{c})(\delta k_{x}^{2}+\delta k_{y}^{2})}$($\delta
k_{x,y,z}$\ are small displacement from $\mathbf{k}_{c}$), which is a linear
dispersion and suggests in neighbourhood of $\mathbf{k}_{c}$, our effective
Hamiltonian is nothing but 3D anisotropic\ massless Dirac fermions.

\begin{figure}[tbh]
\centering\includegraphics[scale=0.5]{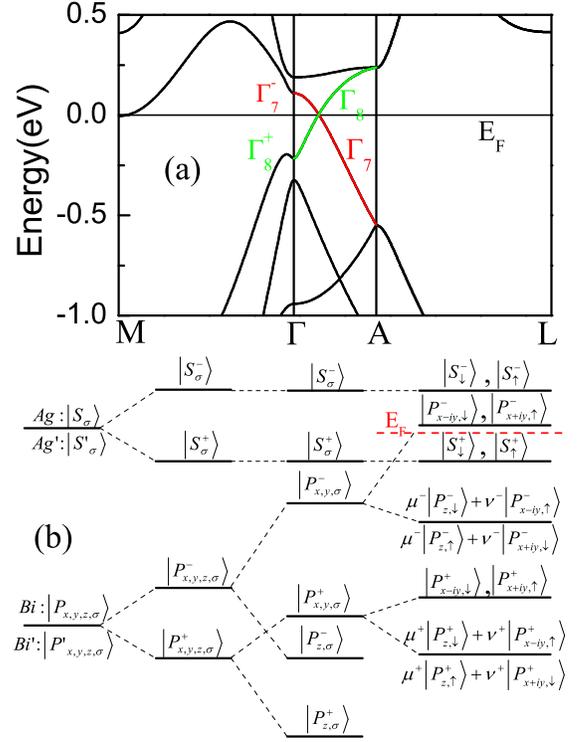}
\caption{(a) Electronic structure of BaAgBi, Green and red line highlights
the different irreducible representation along $\Gamma -A$. (b) Band
evolution around Fermi energy of BaAgBi at $\Gamma $ point, red dashed line
stands for the Fermi energy.}
\end{figure}

We also investigate the BaAgBi and BaCuBi. The electronic properties of
BaCuBi are very similar with that of BaAgBi, we thus only discuss BaAgBi\ at
following. As shown in Fig.3(b), sharply contrast to the Au-6\textit{s}
state in BaAuBi, the Ag-5\textit{s} orbital in BaAgBi is higher in energy
than Bi-6\textit{p} state, consequently the states closed to the Fermi level
become $|P_{x+iy,\uparrow }^{-}\rangle ,\ |P_{x-iy,\downarrow }^{-}\rangle $%
\ and $|S_{\uparrow }^{+}\rangle ,\ |S_{\downarrow }^{+}\rangle $. Similar
with the case in Na$_{3}$Bi\cite{Na3Bi}, due to the strong SOC of Bi-6%
\textit{p}, the $|P_{x+iy,\uparrow }^{-}\rangle $ and$\ |P_{x-iy,\downarrow
}^{-}\rangle $\ states will be pushed up, which results in the band
inversion at $\Gamma $ point. This inversion is confirmed by our DFT
calculation, as shown in Fig.3(a), at the $\Gamma $\ point, the $%
|P_{x+iy,\uparrow }^{-}\rangle ,\ |P_{x-iy,\downarrow }^{-}\rangle $\ is
higher than $|S_{\uparrow }^{+}\rangle ,\ |S_{\downarrow }^{+}\rangle $ by
about 0.34 eV. Along $\Gamma -A$\ line, these two bands belongs to different
($\Gamma _{7}$ and $\Gamma _{8}$) representations, thus there is also a
unavoidable crossing point located at $\Gamma -A$\ line. It is also easy to
prove that around the band touching points, the band dispersion is linear,
thus the crossing points are the Dirac points.

It is well known that breaking the time reversal symmetry will split the
Dirac point into two Weyl points. This family of intermetallic compound with
hexagonal structure indeed has several members with magnetic ion Eu: Eu$XY$ (%
$X$=Cu, Au, Ag; $Y$=As, Sb, Bi) \cite{BaXBi structure,EuCuAs,Eu-Compound}.
Experiments confirm that some of them indeed possess long-range magnetic
configuration \cite{Eu-Compound}. Unfortunately, the Eu$^{2+}$ spins align
ferromagnetically with the \textit{ab} plane, but antiferromagnetically
along the \textit{c}-axis\cite{EuCuAs}, therefore the exchange field is
exactly cancelled at the $XY$-plane of Eu$XY$. Thus breaking the time
reversal symmetry by this type of antiferromagnetic configuration cannot
split the Dirac points, and the compounds of Eu$XY$ have no chance to become
WSM.

\begin{figure}[tbph]
\centering\includegraphics[scale=0.5]{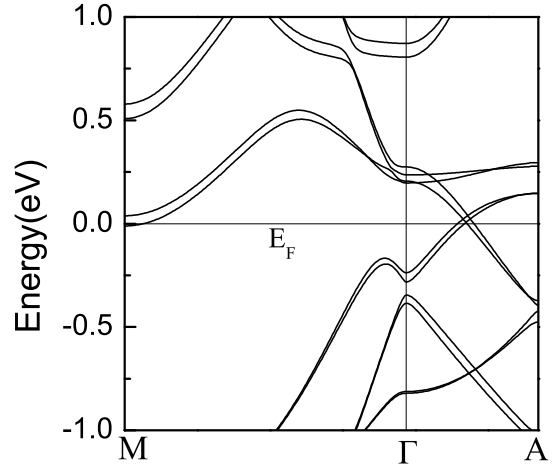}
\caption{Calculated band structure of Eu$_{0.5}$Ba$_{0.5}$AgBi}
\end{figure}

We, however, expect that substituting part of Eu ions by Ba ions, the two
antiferromagnetically coupled Eu plane in Eu$_{x}$Ba$_{1-x}$AgBi may not
exactly cancel each other, and then there is a chance the compound becomes
WSM. To confirm this expectation, we then performed another calculation on Eu%
$_{0.5}$Ba$_{0.5}$AgBi, in which we replace one of the two
antiferromagnetically coupled Eu plane in the unit cell by Ba. As shown in
Fig.4, the Dirac point indeed splits into two Weyl points, and Eu$_{0.5}$Ba$%
_{0.5}$AgBi becomes WSM. We argue that such structure can be typically
synthesized using the cutting-edge film growth technique like MBE. Therefore
we believe that WSM can be realized in the above discussed structures.

In summary, based on density-functional calculation and effective model
analysis, we propose that the Ba$Y$Bi ($Y$=Au, Ag and Cu) are 3D Dirac
semimetals. The nontrivial topological feature is due to \textit{p-p}
inversion for BaAuBi and \textit{s-p} band inversion for BaAgBi and BaCuBi,
and their Dirac points are protected by the $C_{3}$ rotation symmetry. Their
magnetic cousins, i.e Eu$Y$Bi ($Y$=Au, Ag and Cu) are not Weyl semimetals.
However, partial substitution of Eu with Ba ions in Eu$Y$Bi could result in
the Weyl semimetal. Furthermore, our numerical calculation also confirm that
a uniaxial strain along \textit{a}-axis, which breaks the $C_{3}$ rotation
symmetry, will drive BaAgBi into topological insulator.

$Note$. When finalizing our work, we became aware of a recent study by
Borisenko \textit{et al.} \cite{Cava-2}, in which the authors also predict
BaAgBi is a possible 3D DSM, agreeing with our conclusion.

The work was supported by the National Key Project for Basic Research of
China (Grants No. 2011CB922101, 2014CB921104), NSFC under Grants No.
91122035, 11174124 , 11374147 and 61125403. The project is also funded by
Priority Academic Program Development of Jiangsu Higher Education
Institutions.

\section{APPENDIX}

\subsection{Effective Hamiltonian for BaAuBi}

\begin{table*}[tbh]
\centering%
\begin{tabular}{ccc}
\hline
$\Gamma $ matrices & representation & T \\ \hline
$\Gamma _{0}$,$\Gamma _{5}$ & $R_{1}$ & + \\ 
$\{\Gamma _{1},\Gamma _{2}\}$ & $R_{14}$ & + \\ 
$\{\Gamma _{3},\Gamma _{4}\}$ & $R_{15}$ & + \\ 
$\Gamma _{12}$,$\Gamma _{34}$ & $R_{2}$ & - \\ 
$\Gamma _{14}+\Gamma _{23}$ & $R_{3}$ & - \\ 
$\Gamma _{13}-\Gamma _{24}$ & $R_{4}$ & - \\ 
$\{\Gamma _{13}+\Gamma _{24},\Gamma _{14}-\Gamma _{23}\}$ & $R_{6}$ & - \\ 
$\{\Gamma _{15},\Gamma _{25}\}$ & $R_{14}$ & - \\ 
$\{\Gamma _{35},\Gamma _{45}\}$ & $R_{15}$ & - \\ \hline
$d(k)$ & representation & T \\ \hline
$C$, $k_{z}^{2}$, $k_{x}^{2}+k_{y}^{2}$ & $R_{1}$ & + \\ 
$\{k_{x}k_{y}k_{z},\frac{1}{2}(k_{x}^{2}-k_{y}^{2})k_{z}\}$ & $R_{14}$ & -
\\ 
$\{k_{x},k_{y}\},\{(k_{x}^{2}+k_{y}^{2})k_{x},(k_{x}^{2}+k_{y}^{2})k_{y}\},%
\{k_{z}^{2}k_{x},k_{z}^{2}k_{y}\}$ & $R_{15}$ & - \\ 
$k_{z}$, $(k_{x}^{2}+k_{y}^{2})k_{z}$, $k_{z}^{3}$ & $R_{11}$ & - \\ 
$\{k_{x}^{2}-k_{y}^{2},k_{x}k_{y}\}$ & $R_{5}$ & + \\ 
$\{k_{x}k_{z},k_{y}k_{z}\}$ & $R_{6}$ & + \\ 
$k_{x}^{3}-3k_{x}k_{y}^{2}$ & $R_{12}$ & - \\ 
$k_{y}^{3}-3k_{x}^{2}k_{y}$ & $R_{13}$ & - \\ \hline
\end{tabular}%
\caption{The character table of Dirac $\Gamma $ \ matrices and the
polynomials of the momentum $k$ for BaAuBi.}
\end{table*}

The conduction and valence bands of BaAuBi are mainly contributed by four
states: $\left\vert P_{x+iy,\uparrow }^{+}\right\rangle $, $\mu
^{-}\left\vert P_{z,\downarrow }^{-}\right\rangle +\upsilon ^{-}\left\vert
P_{x-iy,\uparrow }^{-}\right\rangle $, $\left\vert P_{x-iy,\downarrow
}^{+}\right\rangle $ and $\mu ^{-}\left\vert P_{z,\uparrow
}^{-}\right\rangle +\upsilon ^{-}\left\vert P_{x+iy,\downarrow
}^{-}\right\rangle $, we thus use these states as the basis to build the
effective model Hamiltonian at the $\Gamma $\ point of BZ. As a $4\times 4$
hermitian matrix, the effective Hamiltonian can be written as $H=\epsilon (%
\mathbf{k})\mathbf{I}+\sum\limits_{i}d_{i}(\mathbf{k})\Gamma
_{i}+\sum\limits_{ij}d_{ij}(\mathbf{k})\Gamma _{ij}$, where $\mathbf{I}$ is
the $4\times 4$ identity matrix, $\Gamma _{i}$\ and $\Gamma _{ij}$ are Dirac
matrices, $\epsilon (\mathbf{k})$, $d_{i}(\mathbf{k})$, and $d_{ij}(\mathbf{k%
})$ are function of momentum \textit{k}.

The Hamiltonian should be invariant under the operation of crystal symmtery
and time reversal symmtery. This requires the function $d_{i}(\mathbf{k})$ [$%
d_{ij}(\mathbf{k})$] and the associated $\Gamma _{i}$ [$\Gamma _{ij}$]
matrices belong to the same irreducible representation. Thus the key problem
is to determine the irreducible representation for $d_{i}(\mathbf{k})$ [$%
d_{ij}(\mathbf{k})$] and $\Gamma $ matrices, which can be done by the
projection-operator method.

The Dirac $\Gamma $ matrices can be written as $\Gamma _{1}=\sigma
_{1}\otimes \tau _{1}$, $\Gamma _{2}=\sigma _{2}\otimes \tau _{1}$, $\Gamma
_{3}=\sigma _{3}\otimes \tau _{1}$, $\Gamma _{4}=\sigma _{0}\otimes \tau
_{2} $, $\Gamma _{5}=\sigma _{0}\otimes \tau _{3}$, and $\Gamma _{ab}=\left[
\Gamma _{a},\Gamma _{b}\right] /2i$\cite{C-X Liu}. The projection operator
is defined as $p^{i}=\frac{l_{i}}{g}\sum\limits_{R\in G}\chi ^{i}(R)P_{R}$,
where $g$ is the group order, $l_{i}$ is the dimension of the $i$th
representation, $R$ denotes the group element i.e. the symmetry operation, $%
\chi ^{i}(R)$ represent the character of group element $R$ in $i$th
representation, $P_{R}$ is the operator of group element $R$.

The double group of D$_{6h}^{4}$\ has 18 classes, and their irreducible
representations are denoted as $R_{1}$ to $R_{18}$ \cite{group book}, and
its character table can be found in Ref\cite{group book}. Based on the basis
mentioned above, one can easily work out the transformation matrix $D_{R}\ $%
for symmetry operator $P_{R}$, which allow us to apply the projection
operator $p^{i}$ on $\Gamma _{a}$: $p^{i}\Gamma _{a}=\frac{l_{i}}{g}%
\sum\limits_{R\in G}\chi ^{i}(R)D_{R}\Gamma _{a}D_{R}^{-1}$, consequently
determine the irreducible representation of $\Gamma _{a}$. Using the same
process, one can also determine the irreducible representation for the
polynomials of $\mathbf{k}$ up to $O(\mathbf{k}^{3})$. We present the
irreducible representation of Dirac $\Gamma $\ matrices and polynomials of $%
\mathbf{k}$, and their transformation under time reversal in Table I.

With the Table I, the effective model Hamiltonian of BaAuBi can be easily
expressed as: $H=\varepsilon _{0}(\mathbf{k})+M(\mathbf{k})\Gamma _{5}+A(%
\mathbf{k})(k_{x}\Gamma _{45}+k_{y}\Gamma
_{35})+Bk_{z}((k_{x}^{2}-k_{y}^{2})\Gamma _{25}+2k_{x}k_{y}\Gamma _{15})$,
where $\epsilon _{0}(\mathbf{k}%
)=C_{0}+C_{1}k_{z}^{2}+C_{2}(k_{x}^{2}+k_{y}^{2})$, $M(\mathbf{k}%
)=M_{0}-M_{1}k_{z}^{2}-M_{2}(k_{x}^{2}+k_{y}^{2})$, $A(\mathbf{k}%
)=A_{0}+A_{1}k_{z}^{2}+A_{2}(k_{x}^{2}+k_{y}^{2})$.

\subsection{Effective Hamiltonian for BaAgBi}

\begin{table*}[tbh]
\begin{tabular}{ccc}
\hline
$\Gamma $ matrices & representation & T \\ \hline
$\Gamma _{0}$,$\Gamma _{5}$ & $R_{1}$ & + \\ 
$\{\Gamma _{1},\Gamma _{2}\}$ & $R_{14}$ & - \\ 
$\{\Gamma _{3},\Gamma _{4}\}$ & $R_{15}$ & - \\ 
$\Gamma _{12}$,$\Gamma _{34}$ & $R_{2}$ & - \\ 
$\Gamma _{14}-\Gamma _{23}$ & $R_{3}$ & - \\ 
$\Gamma _{13}+\Gamma _{24}$ & $R_{4}$ & - \\ 
$\{\Gamma _{13}-\Gamma _{24},\Gamma _{14}+\Gamma _{23}\}$ & $R_{6}$ & - \\ 
$\{\Gamma _{15},\Gamma _{25}\}$ & $R_{14}$ & + \\ 
$\{\Gamma _{35},\Gamma _{45}\}$ & $R_{15}$ & + \\ \hline
$d(k)$ & representation & T \\ \hline
$C$, $k_{z}^{2}$, $k_{x}^{2}+k_{y}^{2}$ & $R_{1}$ & + \\ 
$\{k_{x}k_{y}k_{z},\frac{1}{2}(k_{x}^{2}-k_{y}^{2})k_{z}\}$ & $R_{14}$ & -
\\ 
$\{k_{x},k_{y}\},\{(k_{x}^{2}+k_{y}^{2})k_{x},(k_{x}^{2}+k_{y}^{2})k_{y}\},%
\{k_{z}^{2}k_{x},k_{z}^{2}k_{y}\}$ & $R_{15}$ & - \\ 
$k_{z}$, $(k_{x}^{2}+k_{y}^{2})k_{z}$, $k_{z}^{3}$ & $R_{11}$ & - \\ 
$\{k_{x}^{2}-k_{y}^{2},k_{x}k_{y}\}$ & $R_{5}$ & + \\ 
$\{k_{x}k_{z},k_{y}k_{z}\}$ & $R_{6}$ & + \\ 
$k_{x}^{3}-3k_{x}k_{y}^{2}$ & $R_{12}$ & - \\ 
$k_{y}^{3}-3k_{x}^{2}k_{y}$ & $R_{13}$ & - \\ \hline
\end{tabular}%
\caption{The character table of Dirac matrices and the function $d(k)$ of
BaAgBi.}
\end{table*}

For BaAgBi, the conduction bands are Ag-5\textit{s} states, while the
valence bands are Bi-6\textit{p} states, thus the four basis become $%
\left\vert S_{\uparrow }^{+}\right\rangle $, $\left\vert P_{x+iy,\uparrow
}^{-}\right\rangle $, $\left\vert S_{\downarrow }^{+}\right\rangle $ and $%
\left\vert P_{x-iy,\downarrow }^{-}\right\rangle $. We list the character
table of $\Gamma $ matrices and the function $d(\mathbf{k})$ (expanded as
polynomials of the momentum \textit{k}) and their transformation under time
reversal in Table II. Based on Table II, one can get the effective model
Hamiltonian for BaAgBi: $H=\varepsilon _{0}(\mathbf{k})+M(\mathbf{k})\Gamma
_{5}+A(\mathbf{k})(k_{x}\Gamma _{3}-k_{y}\Gamma
_{4})+Bk_{z}((k_{x}^{2}-k_{y}^{2})\Gamma _{1}+2k_{x}k_{y}\Gamma _{2})$,
where $\epsilon _{0}(\mathbf{k}%
)=C_{0}+C_{1}k_{z}^{2}+C_{2}(k_{x}^{2}+k_{y}^{2})$, $M(\mathbf{k}%
)=M_{0}-M_{1}k_{z}^{2}-M_{2}(k_{x}^{2}+k_{y}^{2})$, $A(\mathbf{k}%
)=A_{0}+A_{1}k_{z}^{2}+A_{2}(k_{x}^{2}+k_{y}^{2})$.

\subsection{Band representation}

\begin{table}[tbph]
\begin{tabular}{ccccccc}
\hline
$D_{6h}^{4}$ & $\Gamma _{7}^{+}$ & $\Gamma _{8}^{+}$ & $\Gamma _{9}^{+}$ & $%
\Gamma _{7}^{-}$ & $\Gamma _{8}^{-}$ & $\Gamma _{9}^{-}$ \\ \hline
$C_{6v}^{1}$ & $\Gamma _{7}$ & $\Gamma _{8}$ & $\Gamma _{9}$ & $\Gamma _{7}$
& $\Gamma _{8}$ & $\Gamma _{9}$ \\ \hline
\end{tabular}%
\label{Table3}
\caption{The compatibility relations between the double group of $C_{6v}^{1}$
and $D_{6h}^{4}$.}
\end{table}

At the $\Gamma $\ point of BZ, each state should belong to an irreducible
representation of the double group of $D_{6h}^{4}$.\ Again, applying the
projection operator onto the conduction and valence states of BaAuBi, we
find that $\left\vert P_{x+iy,\uparrow }^{+}\right\rangle $ and $\left\vert
P_{x-iy,\downarrow }^{+}\right\rangle $ belong to representation $\Gamma
_{7}^{+}$, while $\mu ^{-}\left\vert P_{z,\downarrow }^{-}\right\rangle
+\upsilon ^{-}\left\vert P_{x-iy,\uparrow }^{-}\right\rangle $ and $\mu
^{-}\left\vert P_{z,\uparrow }^{-}\right\rangle +\upsilon ^{-}\left\vert
P_{x+iy,\downarrow }^{-}\right\rangle $ belong to representation $\Gamma
_{8}^{-}$, which had been marked in Fig.2(a). Different from $\Gamma $\
point, the symmetry of $\Gamma -A$ line is $C_{6v}^{1}$. We show the
compatibility relations between the double group of $D_{6h}^{4}$\ and $%
C_{6v}^{1}$\ in Table III. It is clear that the representation $\Gamma
_{7}^{+}$\ and $\Gamma _{8}^{-}$, evolute to $\Gamma _{7}$\ and $\Gamma _{8}$%
, respectively.

For BaAgBi, the valence/conduction states at the $\Gamma $\ point of BZ
belong $\Gamma _{8}^{+}/\Gamma _{7}^{-}$, and will change to $\Gamma _{8}$\
and $\Gamma _{7}$ along $\Gamma -A$ line according to Table III.

\end{document}